\documentclass[usenatbib]{mn2e}
\usepackage{times}
\usepackage{epsfig}
\usepackage{amsmath}
\usepackage{amssymb}
\title[Constraints on intragroup stellar mass from hostless Type Ia
  supernova]{Constraints on intragroup stellar mass from hostless Type Ia supernovae}

\author[McGee \& Balogh]{Sean L. McGee$^{1}$\thanks{Email:
    s2mcgee@uwaterloo.ca} and  Michael L. Balogh$^{1}$
\\
$^{1}$Department of Physics and Astronomy, University of Waterloo, Waterloo, Ontario, N2L 3G1, Canada\\
}

\date{\today}

\def\LCDM{$\Lambda$CDM$~$}
\def\Mdoth{$h^{-1}~$M$_\odot$}
\def\Mpch{$h^{-1}~$Mpc$~$}

\newcommand{\kms}{${\rm km}{\rm s}^{-1}$}
\newcommand{\kmsmpc}{\>{\rm km}\,{\rm s}^{-1}\,{\rm Mpc}^{-1}}
\begin{document}
\maketitle

\begin{abstract}

We probe the diffuse stellar mass in a sample of 1401 low redshift
galaxy groups (10$^{13}$ - 10$^{14}$ \Mdoth) by examining the rate of
hostless Type Ia supernova (SNe Ia) within the groups. We correlate
the sample of confirmed SNe Ia from the SDSS supernova survey with the
positions of our galaxy groups, as well as with the resolved galaxies
within them. We find that 19 of the 59 SNe Ia within the group sample
have no detectable host galaxy, with another three ambiguous instances.
This gives a robust upper limit that a 
maximum of $2.69^{+1.58}_{-1.34}$\% percent of the group's total mass
arises from diffuse stars in the intragroup medium. After correcting
for a contribution from ``prompt'' SNe occurring within galaxies, and
including a contribution from those which arise in dwarf galaxies
below our photometric limit, we find that only 1.32$^{+0.78}_{-0.70}$
$\%$ of the group's total mass is likely in the form of diffuse stellar
mass. Combining this result with the galaxy stellar mass functions of
\citet{yangstellar}, we find that 47$^{+16}_{-15}$ $\%$ of the 
{\it stellar mass} in our groups is in the form of diffuse light, so that
stars make up a fraction 0.028$^{+0.011}_{-0.010}$ of the {\it total} group
mass. Galaxy groups appear to be very efficient in disrupting stellar mass
into a diffuse component; however, stars still make up a small fraction
of the group mass, comparable to that seen in rich clusters.  This
remains a challenge to galaxy formation models.

\end{abstract}


\section{Introduction}

Diffuse light in the halos of galaxy clusters has been extensively
studied since the discovery of 'swarms of stars' between galaxies in
the Coma cluster by \citet{zwicky}. Although puzzling at the time,
diffuse stellar light is now thought to be a natural consequence of
hierarchical structure formation, which causes merging galaxies to
shed some of their stellar material \citep{willman_icl, purcell}. The diffuse light
in individual massive clusters has been the focus of many
observational studies,which find that the fraction of the total
cluster light in this component is 10-30$\%$
\citep[e.g.,][]{thuan, scheick}. By stacking $\sim$ 700
clusters, \citet{zibetti} was able to tightly constrain the diffuse
light within 500 kpc of the cluster center to be 10.9 $\pm$ 5.0 $\%$
of the total cluster light.

At the other extreme, the diffuse light around galaxies like the Milky
Way is similarly thought to arise from the tidal heating and stripping
of infalling dwarf galaxies \citep[e.g.,][]{searle,
bullock,font_stellar}. However, in this regime, the fraction is only
$\sim$ 1-2 $\%$ of the total light in the halo
\citep[e.g.,][]{chiba,yanny,law}.  Probing the intermediate halo mass
regime, that of galaxy groups, is important in order to reconcile
these two extreme regimes.

Galaxy groups are difficult to observe observationally and thus, a
study of the intragroup medium (IGM) in typical systems has not been
done. Studies of the more extreme compact groups have found a large
variation in the group to group diffuse light, ranging from 5$\%$ up
to as much as 45 $\%$ of the total light
\citep{white_light,darocha1,darocha2}.

Galaxy groups have velocity dispersions comparable to that of the most
massive galaxies within them, and as such are a prime location for the
mergers, tidal stripping and shredding which is thought to give rise
to the diffuse light. Thus, the amount of diffuse stellar mass in
galaxy groups is a direct probe of the efficiency with which infalling
galaxies are disrupted. This could have important implications for
semi-analytic models, which produce too many faint red galaxies
\citep{weinmann_semi, gilbank}.

Type Ia supernova (SNe Ia) likely result when a carbon-oxygen white
dwarf acquires additional mass from a companion star, which causes a
thermonuclear explosion. However, the exact origin of the progenitor
material is uncertain, which makes direct prediction of SN Ia rates
from stellar population modeling difficult. It is typical to
parametrize the rate and then measure the coefficients
empirically. The most common model for the SN Ia rate is the so called
A+B model, which assumes that SN Ia arise from two distinct
channels. In this model, there is a ``prompt'' component which depends
on the current star formation rate (SFR) of the host and a ``delayed''
component which traces the host's stellar mass
\citep{scannapieco,graham}. This model was motivated by the
observation that the SN Ia rate is $\sim$ 20-30 times higher in late
type galaxies than in early type galaxies of the same mass
\citep{mannucci}.

Relatively little attention has been given to SNe which may be hosted
by diffuse material within groups and clusters. But, \citet{galyam}
found that 2 of the 7 cluster SNe Ia discovered during a survey of low
redshift Abell clusters were not associated with galaxies. Recently,
\citet{sand} have begun a search of 60 X-ray selected galaxy clusters
with the expectation of finding $\sim$ 10 intracluster SNe Ia. They
identified three intracluster candidates in early data, all of which
were actually outside of R$_{200}$, implying a relative deficit of
intracluster mass at small cluster radii.

This Letter is complementary to these studies, as we constrain the
diffuse stellar mass in relatively low-mass galaxy groups, by
correlating type Ia supernovae with a large group catalogue, to
identify supernovae without a resolved galaxy host. In \textsection
\ref{data}, we discuss our supernova and group sample, as well as
introducing the derived galaxy properties we use. In \textsection
\ref{identify}, we discuss the procedure we use to identify the hosts
or lack of hosts of the supernovae. Finally, in \textsection
\ref{results}, we discuss the constraints this places on the diffuse
stellar mass in galaxy groups. Throughout this paper, we adopt, as was
done during the assembly of the group catalogue, a \LCDM cosmology
with the parameters of the third year WMAP data, namely $\Omega_{\rm
m} = 0.238$, $\Omega_{\Lambda}=0.762$, $\Omega_{\rm b}=0.042$,
$n=0.951$, $h=H_0/(100 \kmsmpc)=0.73$ and $\sigma_8=0.75$
\citep{spergel}.

\section{Data}\label{data}

In this Letter, we require a uniformly selected sample of SNe Ia and a
large sample of galaxy groups over a large area of the sky to find
enough hostless SNe Ia to constrain the diffuse stellar
mass. Therefore, we will examine the region of Stripe 82 in the Sloan
Digital Sky Survey (SDSS) which hosts the SDSS supernova survey.

\subsection{SDSS supernova survey}

The SDSS supernova survey was designed to identify SNe Ia at low
redshift (0.05 $<$ z $<$ 0.35) by imaging an area of 300 sq. degrees
multiple times over a period of 5 years \citep{supernova_tech}. Once
identified, the SNe Ia candidates are spectroscopically followed up and
classified \citep{supernova_follow,supernova_spect}. This has resulted
in the largest collection of supernova at low redshift. Crucially, for
our purposes, the scanning throughout the 300 sq. degrees is
very uniform, resulting in a consistent detection
threshold. For the following, we restrict our analysis to the 368
confirmed SNe Ia within 0.1 $<$ z $<$ 0.2. We note that the SDSS
supernova survey also identifies type II SNe. However, despite
being more numerous, they are are also much fainter \citep{bazin}. Thus
spectroscopic follow-up of type II SNe was limited to those at $z<0.06$,
which is below our redshift limit. 

\subsection{Group catalogue}

Reliable and representative samples of galaxy groups can be found
using friends of friends algorithms in large redshift surveys
\citep[e.g.,][]{huchra,carlberg,berlind}. We use a highly complete group sample
defined by \citet{yanggroups}, who have applied their 'friends of
friends'-like algorithm to the SDSS fourth data release to produce a
sample of $\sim$ 300,000 galaxy groups with masses as low as
10$^{11.5}$ \Mdoth. An important aspect in the construction of this
group catalogue is the subsequent mass estimates, which are obtained by
essentially ranking groups by their total luminosity or total stellar
mass and associating these rankings with the expectations of a \LCDM
halo occupation model. We use ``Sample I'' from \citet{yanggroups}, which
exclusively uses galaxies with SDSS spectroscopic redshifts. We chose
this sample, as opposed to the other samples, which add in existing
redshifts from the literature, because we are principally concerned
with obtaining a uniformly selected population of galaxy groups. In
this Letter, we use the halo mass rankings obtained from the total
group stellar mass and restrict our analysis to groups with
masses greater than 10$^{13}$ \Mdoth, which leaves a final sample of
1401 groups between z=0.1 and z=0.2 within the SDSS supernova legacy
survey area. The galaxy groups have a median halo mass of 1.98
$\times$ 10$^{13}$ \Mdoth and a mean value of 3.86 $\times$ 10$^{13}$
\Mdoth with a standard deviation of 3.16 $\times$ 10$^{13}$
\Mdoth. 

We make use of measurements of the stellar mass and star formation
rates of the galaxies within the main galaxy spectroscopic sample of
the SDSS. Our stellar masses are taken from \citet{kauffmann_mass},
and were determined by comparing large libraries of stellar population
models with line indices and broad band photometry. The star formation
rates we use were measured by \citet{brinchmann_sfr} using principally
H$\alpha$ emission along with continuum properties (e.g., 4000-$\AA\ $
break).

\begin{figure}
\leavevmode \epsfysize=8cm \epsfbox{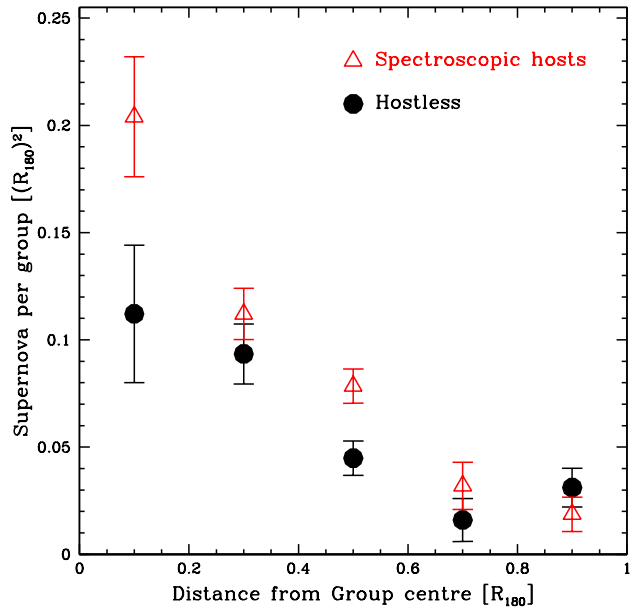}
\caption{The distribution of type Ia supernovae as a function
 of distance from the group centre. The red triangles represent the
 rate of SNe Ia hosted by the sample of spectroscopic galaxies with
 the group, while black circles signify the rate of hostless SNe Ia
 within the group sample. $R_{180}$ corresponds to a distance of 0.79
 \Mpch for a group with the mean mass of our sample ( 3.86 $\times$ 10$^{13}$ \Mdoth).
 }
\label{radi_plot}
\end{figure}

\section{Hosted and hostless group SNe Ia}\label{identify}

Our goal is to obtain a complete sample of SNe Ia which reside within
galaxy groups in our sample. We define group membership as those SNe
Ia which are within the projected group virial radius and which are
$\pm$ 3000 km/s from the group redshift. This velocity range is larger
than the typical velocity dispersion of our groups, but is chosen to be 2
sigma of the precision with which redshifts can be derived from SNe Ia
spectral features \citep[ie. $\sim$ 1500 \kms,][]{supernova_tech,
cooper_sn}. Of the 368 SNe Ia in our complete sample, 59 are matched
to one of our galaxy groups.

We search for host galaxies of the 59 group SNe, by matching to all
group galaxies in the main galaxy sample of SDSS.  We assume a SN is
hosted if it is within a distance equal to twice the size of the
radius enclosing 90$\%$ of the galaxy's light. This radius corresponds
to a physical scale of 9--14 kpc/h for most of the galaxies in the
spectroscopic sample, and is slightly smaller for those in the
photometric sample. However, for a few of the bright group/cluster
galaxies this distance is much larger because of their extended
haloes. For this reason we have capped the search radius to be a
maximum of 40 kpc/h.  In practise, this changes the membership of just
one SNe, which is 78 kpc/h from a BCG. By matching within our search
radius and a velocity window of $\pm$ 3000 km/s, we find that 23 of
the 59 group SNe are hosted by galaxies within the main spectroscopic
galaxy sample. The spectroscopic galaxy sample only reaches a
magnitude limit of $r=17.77$, so we must extend our search for the
hosts of the remaining SNe to the photometric catalogues, which reach
95 $\%$ completeness at $r$ $\sim$ 22.2. This is an absolute magnitude
of M$_r$$\approx -15$ to $-16$ in our sample. We search for hosts of
the remaining SNe in the photometric catalogues again within our
scaled radius, and now use the photometric redshifts supplied by the
SDSS pipeline. Fourteen of the remaining 36 SNe were matched with host
galaxies using this criteria. Finally, we relax the requirement that
the galaxy photometric redshift is within the range of the SN
spectroscopic redshift, and find an additional three galaxies are
matched. This leaves us with a sample of 19 SNe with no host galaxy in
the photometric catalogue. The images of these SNe, as with the entire
sample, were visually inspected to confirm that they appeared as truly
hostless SNe. Given the limit of our photometric catalog, we might
expect that some of these 19 supernovae are actually hosted by
galaxies which are below our detection threshold. Integrating the
$r$-band luminosity function of SDSS at z $\sim$ 0.1
\citep{blanton_lstar}, only 3 $\%$ of the total galaxy luminosity is
below our detection limit. If we adopt a luminosity function typical
of rich galaxy clusters, which have a steep faint end slope
\citep{milne}, as much as 5 $\%$ of the total galaxy luminosity may go
undetected in our images. Thus, if the number of supernova scale with
luminosity, then it is likely that $\sim 2\pm 1$ of the apparently
hostless galaxies are actually hosted by very faint galaxies (all
uncertainties are $1 \sigma$). In summary, we find that of the 59
total group SNe, 23 are matched to spectroscopically confirmed galaxy,
14 are matched to galaxy via photometric redshift, and a further 3
appear associated with other galaxies. This leaves a sample of 19
apparently hostless SNe.

\section{Results} \label{results}

We are first interested in the distribution of the SNe Ia sample
within our galaxy groups. This is presented in Figure \ref{radi_plot},
in which we show the SN Ia per area per group as a function of
groupcentric distance in units of R$_{180}$ as calculated in
\citet{yanggroups}. The SN Ia counts per area decline steeply with
groupcentric distance, which indicate they are associated with the
galaxy group. Also shown is the rate of SN Ia hosted by group galaxies
within the spectroscopic sample. Given the small numbers and
uncertainty on each point, the distributions are similar. There is no
evidence that the hostless SN, and thus the diffuse stellar mass, are
distributed differently from the galaxy population.

Assuming that all 19 apparently hostless galaxies are truly hostless,
and making the simplifying assumption that the stellar population in
the intragroup medium is the same as within the group galaxies, our
measurement immediately tells us the diffuse stellar mass represents
$\sim 32^{+4}_{-3}$ per cent of the total stellar mass in these
groups, where the quoted statistical uncertainty is the $1\sigma$
confidence limit.  We now explore the systematic uncertainties and
consequences associated with this measurement.

\subsection{Upper limit on the diffuse group stellar mass}

As discussed above, without a definite model of the origin of SNe Ia,
the conversion of a SNe number count into the underlying stellar mass
is uncertain. Therefore, we first try to derive the ${\it maximum}$
diffuse stellar mass possible given the population of hostless SNe
Ia. First, we will assume that the 22 supernova without a
spectroscopic or photo-z confirmed host are true descendants of diffuse
group stellar mass. This includes the 19 supernova with no apparent
host as well as three which appear to be located within galaxies, but
whose galaxies have photometric redshifts inconsistant with both the
SN redshift and the redshift of the group. To obtain a robust upper
limit we here further assume that there is no contribution from faint,
undetected hosts. It is worth pointing out that this is also an upper
limit because it is much easier to detect SNe Ia in hostless
environments, where the contrast is high, than it is for those which
occur in galaxies. We will assume that the remaining 37 supernova
occur within galaxies.

We now need to derive a relation between the stellar mass and the
number of resulting ``delayed'' supernova. To do this, we must
accurately know the stellar mass and star formation rate of a
population of galaxies. We will use our sample of galaxies which are
spectroscopically confirmed and which are within our groups. Using the
stellar masses of \citet{kauffmann_mass} we find that the
spectroscopic sample of galaxies in our 1401 groups has a total
stellar mass of 6.55 $\times$ 10$^{14}$ \Mdoth and, using the star
formation rates of \citet{brinchmann_sfr}, the total star formation
rate is 2.62 $\times$ 10$^{4}$ \Mdoth/yr. We can now use the empirical
relation of \citet{dilday}, which was derived from the SDSS supernova
survey, to find out how many of the 23 SNe in the spectroscopic sample
are ``prompt'' SN and how many are ``delayed''. \citeauthor{dilday}
finds that the supernova rate, r, is given by,
r=A$\rho$+B$\dot{\rho}$, where $\rho$ is the stellar mass,
$\dot{\rho}$ is the star formation rate, A=2.8$\pm$1.2 $\times$
10$^{-14}$ SNe M$^{-1}_\odot$ yr$^{-1}$ and B=9.3$^{+3.4}_{-3.1}$
$\times$ 10$^{-4}$ SNe M$^{-1}_\odot$. Using this formula, we find
that 43$^{+18}_{-16}$ $\%$ of 23 SNe are ``delayed'' and 57
$^{+16}_{-18}$ $\%$ are ``prompt''. So, we see that there are
9.9$^{+4.1}_{-3.9}$ ``delayed'' SNe (=43$^{+18}_{-16}$ \% $\times$23),
which arise from an underlying stellar mass of 6.55 $\times$ 10$^{14}$
\Mdoth. Thus, we have the relation that 1 delayed SN arises from an
underlying stellar mass of 6.62$^{+3.92}_{-3.54}$ $\times$ 10$^{13}$
\Mdoth. We will us this relation throughout the rest of the paper to
relate the stellar mass to a quantity of delayed SNe.

\begin{figure}
\leavevmode \epsfysize=8cm \epsfbox{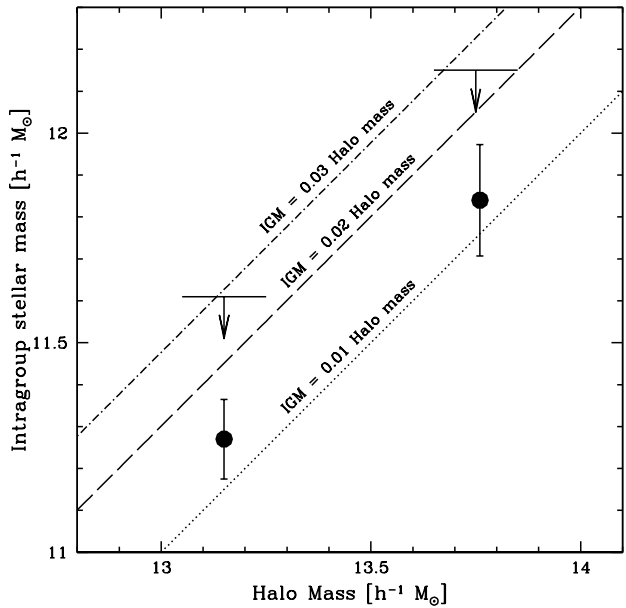}
\caption{The measurements of intragroup stellar mass for our sample,
  assuming no prompt component of SN. The thick black line with the
  arrow represent our upper limits, while the dots are the best
  estimate after accounting for a contribution from possible host
  galaxies below the survey magnitude limit. Also shown are three
  equality lines where the intragroup stellar mass equals 0.03
  (dot-dashed line), 0.02 (dashed line) or 0.01 (dotted line) of the
  total halo mass.  }
\label{mass_plot}
\end{figure}

The diffuse stellar mass is expected to be relatively old, as the
intragroup mass is not expected to be able to form stars in-situ. Any
star formation which results from the stripping of gas from galaxies
will occur near the galaxy, and likely would appear as if it was
hosted by that galaxy \citep{sun, sun2}. Indeed, the intracluster
light observed to date has been universally old \citep{zibetti}. So,
it is reasonable to assume that there exist no ``prompt'' supernova in
the intragroup mass, particularly since we are here after a robust
upper limit on the diffuse stellar mass.  Thus, we can conclude that
the diffuse component hosts at most 22 ``delayed'' SN, and we can use
the relation found above that 1 ``delayed'' SN arises from a stellar
mass of 6.62$^{+3.92}_{-3.54}$ $\times$ 10$^{13}$ \Mdoth, to see that
the total stellar mass in the diffuse component is $\leq
1.46^{+0.86}_{-0.73}$ $\times$ 10$^{15}$ \Mdoth. Therefore, the total
diffuse stellar mass per group is $\leq
1.04^{+0.61}_{-0.52}\times10^{12}$\Mdoth, or $\leq
2.69^{+1.58}_{-1.34}$\% of the halo mass, given the average halo mass
per group of 3.86 $\times$ 10$^{13}$ \Mdoth.  Taking the upper limit
allowed by the statistical uncertainty, our robust upper limit on the
fraction of total mass that occurs in the diffuse stellar component is
$4.3$\%.  In Figure \ref{mass_plot}, we show these upper limits in two
bins of halo mass.  We discuss the implications of this in
Section~\ref{sec-discuss}.

\subsection{Best estimate of the intragroup mass}

We have presented a robust upper limit on the quantity of intragroup
stellar mass in galaxy groups.  We now attempt to make a more
realistic calculation of the contribution from undetected host
galaxies, to arrive at an estimate of the actual mass.

In the previous section, we assumed all 22 hostless SN were actually
associated with the diffuse mass. However, three of these appear to be
associated with photometric galaxies, which have inconsistent
photometric redshifts.  From their proximity to these galaxies we
think it is more likely that the photometric redshift is incorrect,
and these are really ``hosted'' SN, within the group.  Moreover, we
have argued that for typical groups up to 5 $\%$ of the group galaxy
light is expected to be in unresolved hosts. Therefore, if stellar
light scales as stellar mass, we would expect $\sim$ 2 of the remaining
19 apparently hostless SNe to be ``delayed'' SNe hosted in unresolved
galaxies.

However, while the ``delayed'' SN rate should only depend on the total
stellar mass, galaxies which are strongly star forming will have a
greater ``prompt'' component than galaxies within our spectroscopic
sample. To estimate this effect, we must adopt a scaling between the
stellar mass and star formation rate. As low mass galaxies are usually
uniformly star forming, we assume that the low mass galaxies double
their stellar mass in a Hubble time
(e.g. SFR=$\frac{M/M_\odot}{10^{10} yr}$). Using the relation of
\citeauthor{dilday}, we find that the number of prompt SN is equal to
$9.3/2.8=3.3$ times the number of delayed SN, so that $\sim$ 6 of the
hostless SNe are likely due to ``prompt'' explosions in unresolved
galaxies. Taken together, this means only $\sim$ 11 of our hostless
SNe are likely true intragroup SNe. Again, we assume that all true
intragroup SNe are ``delayed'' SNe, because the intragroup mass is
uniformly old. Therefore, we can use the previous scaling which stated
that 1 ``delayed'' SN arises from a stellar mass of
6.62$^{+3.92}_{-3.54}$ $\times$ 10$^{13}$ \Mdoth, to find that a total
mass of 7.28$^{+4.31}_{-3.89}$ $\times$ 10$^{14}$ \Mdoth is in the
diffuse component.  On average, this corresponds to
$5.20^{+3.08}_{-2.78}$ $\times$ 10$^{11}$ \Mdoth per group,
representing a fraction 1.32$^{+0.78}_{-0.70}$ \% of the total halo
mass.  The black points in Figure \ref{mass_plot} shows the final best
estimate of the diffuse stellar mass in two bins of halo mass, after
all the corrections discussed above. The diffuse stellar fraction is
statistically indistinguishable in the two mass bins, with $\sim
1.3^{+0.73}_{-0.65}$ $\%$ in the lowest mass groups, and
$1.2^{+0.62}_{-0.56}$ $\%$ for the more massive systems.

\begin{figure}
\leavevmode \epsfysize=8cm \epsfbox{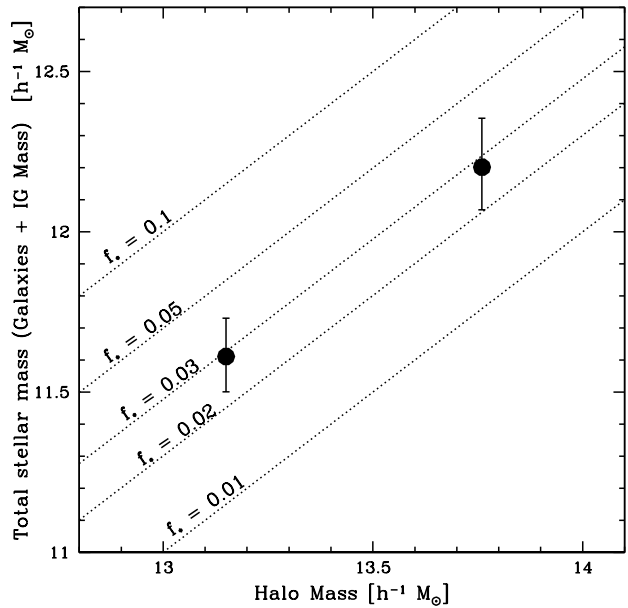}
\caption{Total stellar mass in our sample of groups. This includes the
 diffuse intragroup mass measured in this paper added to the stellar
 mass of galaxies in the same sample as determined by
 \citet{yangstellar}. The dashed lines represent lines of constant
 total stellar fraction,
 f$_*$=$\frac{\rm{Total~Stellar~Mass}}{\rm{Total~Halo~ Mass}}$. Our
 sample indicates the stellar fraction these groups is $\sim$
 0.02-0.04.  }
\label{totmass_plot}
\end{figure}

A potential systematic error in our results could occur if the SDSS
supernova pipeline and/or spectroscopic follow-up is biased towards
finding SNe without hosts. The first issue, that of the finding
algorithm, was explored by \citet{dilday}, who found that the pipeline
uncovers $>$ 98 $\%$ of SNe Ia to a redshift of 0.2.  Even if the
completeness were only 95$\%$ and entirely biased toward hostless SN
(i.e. assuming the 5$\%$ missed SN are associated with galaxies), our
result on the amount of mass in diffuse form would only change from
1.32$\%$ to 1.17$\%$.  Regarding the spectroscopic follow-up,
\citet{supernova_follow} state that essentially all of the z$<$0.15
type Ia SNe were spectroscopically followed up, and our results are
unchanged if we restrict just to this sample.  Thus we conclude that
neither of these selection affects have a significant affect on our
results.

\section{Discussion and Conclusions}\label{sec-discuss}

We have measured the diffuse stellar mass in galaxy groups using the
rate of hostless SNe Ia. We find that 1.32$^{+0.78}_{-0.70}$ $\%$ of
the total halo mass of a group is in the form of this diffuse
intragroup stellar mass. Although many numerical predictions exist for
the mass of the diffuse intragroup material, they are usually stated
as a fraction of the total stellar mass. To obtain this measure we
will use the stellar mass functions of \citet{yangstellar}, which were
measured for the same sample of groups we use. They find that the
stellar mass function of the group galaxies is well approximated by a
log-normal distribution for the central galaxy and a modified
Schechter function for the satellite population. Integrating these
functions over all galaxies reveals that the fraction of stars in
galaxies is $\sim 1.5$ \% of the halo mass for groups in our mass
range\footnote{A very similar result is obtained if we use the total
stellar mass within the spectroscopic sample, and include a small
correction for the stellar mass below the spectroscopic
limit.}. Therefore, $\sim$ 47$^{+16}_{-15}$ $\%$ of the stellar mass
in our groups is in the diffuse component.This is significantly higher
than the estimates of \citet{purcell}, who use Press-Schechter
formalism along with analytic models for the disruption and stripping
of halos to estimate that $\sim$ 20 $\%$ of the stellar mass is in a
diffuse form. It is also significantly larger than that seen in the
most massive clusters, which is more typically about 10$\%$
\citep{zibetti}.

In Figure \ref{totmass_plot}, we show the total stellar mass as a
function of halo mass for our sample. This shows that, with the
uncertainty, the total stellar fraction in our groups is
0.028$^{+0.011}_{-0.010}$, with a robust upper limit of $0.058$.  We
find some tension with a previous measurement by \citet{baryon_cens}
of the stellar fraction in a few similar sized groups. They found
$\sim$ 6$\%$ of the total halo mass was in some form of stars in these
sized groups, while massive clusters have stellar fractions of only
$\sim$ 0.02.  Our robust upper limit on the stellar fraction is
actually just consistent with their measurement, but this conservative
limit would imply that there is almost three times as much stellar
mass in the IGM as in the galaxies; not only is this almost certainly
unreasonable, it is also much higher than \citet{baryon_cens}
themselves find.  In fact, the fraction of stellar mass we find in the
IGM ($\sim 47$ per cent) is in reasonable agreement with their study;
the difference in our result is in the stellar mass fractions
themselves, as evident in \citet{yangstellar}.  Since it is expected
that clusters are formed from the buildup of groups \citep{berrier,
mcgee_accretion} and given that mergers do not destroy stellar mass,
\citet{balogh_buildup} showed that a group stellar fraction as high as
found by \citet{baryon_cens} was incompatible with the low stellar
fractions in clusters, and argued that the halo masses of the
\citet{baryon_cens} groups were underestimated. The lower measurement
of the group stellar fraction from this work and \citet{yangstellar}
supports this conclusion.

We conclude, therefore, that despite the significant contribution from
intragroup stars, the stellar mass fraction in groups is not
significantly larger than in massive clusters.  This suggests that the
low gas fraction in groups \citep[e.g.][]{Vik+06,Sun+09} cannot likely
be explained via an extraordinarily high star formation efficiency.



\bibliography{ms1}

\end{document}